\documentclass[twocolumn,amsmath,amssymb]{revtex4}
\usepackage{graphicx}
\usepackage{dcolumn}
\usepackage{bm}
\begin{document}

\title{Gap Structure of the Hofstadter System of Interacting Dirac 
Fermions in Graphene}
\author{Vadim M. Apalkov}
\affiliation{Department of Physics and Astronomy, Georgia State University,
Atlanta, Georgia 30303, USA}
\author{Tapash Chakraborty$^\ddag$}
\affiliation{Department of Physics and Astronomy,
University of Manitoba, Winnipeg, Canada R3T 2N2}

\date{\today}
\begin{abstract}
The effects of mutual Coulomb interactions between Dirac fermions in monolayer 
graphene on the Hofstadter energy spectrum have been investigated. For two flux 
quanta per unit cell of the periodic potential, interactions open a gap in each 
Landau level with the smallest gap in the $n=1$ Landau level. For more flux quanta 
though the unit cell, where the noninteracting energy spectra have many gaps in 
each Landau level, interactions enhance the low-energy gaps and strongly 
suppress the high-energy gaps and almost closes a high-energy gap for $n=1$. The 
signature of the interaction effects in the Hofstadter system can be probed 
through magnetization which is governed by the mixing of the Landau levels 
and is enhanced by the Coulomb interaction.  
\end{abstract}
\maketitle

The dynamics of an electron in a periodic potential subjected to a perpendicular
magnetic field has a long history \cite{review,harper,langbein,hofstadter}. 
Hofstadter's numerical solution of the Harper equation \cite{harper} in the 
tight-binding model demonstrated in 1976 that \cite{hofstadter} the magnetic 
field splits the Bloch bands into subbands and gaps. The resulting energy 
spectrum, when plotted as a function of the magnetic flux per lattice cell 
reveals a fractal pattern that is known in the literature as Hofstadter's 
butterfly (due to the pattern resembling the butterflies). A few 
experimental efforts to detect the butterflies have been reported in the
literature. The earlier ones involved artificial lateral superlattices on 
semiconductor nanostructures \cite{geisler_04,albrecht_01,ensslin_96}, more 
precisely antidot lattice structures with periods of $\sim$100 nm. 
The large period (as opposed to those in natural crystals) of the 
artificial superlattices helps to keep the magnetic field in a reasonable range 
of values to observe the fractal pattern. Measurements of the quantized Hall 
conductance in such a structure indicated, albeit indirectly, the complex pattern
of gaps that were expected in the butterfly spectrum. Hofstadter butterfly 
patterns were also predicted to occur in other totally unrelated systems, such 
as, propagation of microwaves through a waveguide with a periodic array of 
scatterers \cite{microwave} or more recently, with ultracold 
atoms in optical lattices \cite{optical_lattice}.

Dirac fermions in monolayer and bilayer graphene \cite{david_PRL,aoki,abergeletal} have 
been found to be the most promising objects thus far, where the signature of the 
recursive pattern of the Hofstadter butterfly has been unambiguously reported 
\cite{dean_13,hunt_13,geim_13}. Here the periodic lattice with a period of 
$\sim$ 10 nm was created by the Moire pattern that appears when graphene is 
placed on hexagonal boron nitride with a twist \cite{moire_1,moire_2}. Although 
the period here is much shorter than that in semiconductor nanostructures, the 
unique properties of graphene helps to create a robust butterfly pattern 
\cite{note_1}. Theoretical studies of the butterfly pattern in monolayer \cite{rhim} 
and bilayer graphene \cite{nemec} systems were also reported earlier. 

In comparison to the numerous studies of noninteracting fermions in the 
butterfly problem, there are very few papers that report on the effects of 
electron-electron interactions on the fractal energy spectra. Hartree \cite{vidar} 
or mean-field approaches, reported earlier in conventional two-dimensional 
electron systems \cite{read,doh_salk} indicated that although the butterfly 
pattern remains intact, additional gap structures are generated by the Coulomb 
interaction. The unique magnetic properties of Dirac fermions in graphene 
\cite{aoki,abergeletal,ssc_13} however provides a new frontier for exploration of 
the intricate structure of the magnetic butterflies. As stated above, graphene 
seems to be the best system to observe the fractal energy spectrum. Therefore, 
it is important to understand the role interacting Dirac fermions play in the 
Hofstadter spectrum in graphene. Here we present our studies of the gap structure 
in the energy spectra due to the Coulomb interaction between Dirac fermions 
in monolayer graphene. Coulomb interaction in the presence of a strong magnetic 
field plays an important role in monolayer and bilayer graphene 
\cite{david_PRL,aoki,abergeletal,ssc_13,mono_FQHE,bi_FQHE}. However, it is quite a challenging 
task to evaluate the role of Coulomb interactions in the present system due to 
the complexities of the Hofstadter energy spectra. Our studies indicate that 
the influence of the Coulomb interaction on the energy gap is highly nontrivial 
in this case. This is also reflected in the magnetization in the 
Hofstadter model of graphene. 

\begin{figure}
\begin{center}\includegraphics[width=6cm]{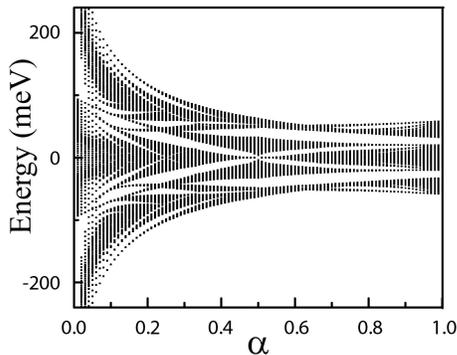}\end{center}
\caption{Single-electron energy spectra of Dirac fermions in graphene in a
magnetic field and a periodic potential $V(x,y)$ with period $a^{}_0 = 20$ nm
and amplitude $V^{}_0 = 35$ meV. The energy
spectra are shown as a function of the number of flux quantum per unit cell. The
results are for the $n=0$ and $n=\pm 1$ Landau levels (LLs).
}
\label{fig1}
\end{figure}

We consider a monolayer graphene in a periodic external potential that has 
the following form  
\begin{equation}
V(x,y) = V^{}_0 \left[\cos(q^{}_x x)+\cos(q^{}_y y)\right],
\label{Vxy}
\end{equation}
where $V^{}_0$ is the amplitude of the periodic potential, $q^{}_x = q^{}_y = 
q^{}_0 = 2\pi/a^{}_0$, and $a^{}_0$ is a period of the external potential 
$V(x,y)$. We begin with the single-particle energy spectrum of the electron 
system in the periodic potential (\ref{Vxy}) and in an external perpendicular 
magnetic field, $B$. The corresponding Hamiltonian then is
\begin{equation}
{\cal H} = {\cal H}^{}_B  + V(x,y),
\label{HV}
\end{equation}
where  ${\cal H}^{}_B$ is the Hamiltonian of an electron in graphene in a 
perpendicular magnetic field. The electron energy spectrum of graphene in a 
magnetic field has two-fold spin and two-fold valley degeneracy. This degeneracy
cannot be lifted by the periodic potential. In that case, for the 
single-electron system we consider the spectrum for a given valley, say valley 
$K$, and a given component of the spin. The corresponding Hamiltonian 
${\cal H}^{}_B$ can now be written \cite{aoki,abergeletal,ssc_13} 
\begin{equation}
{\cal H}^{}_B = \frac{\gamma}{\hbar } \left( 
\begin{array}{cc}
    0 & \pi^{}_x - i \pi^{}_y   \\
    \pi^{}_x + i \pi^{}_y & 0 
\end{array} 
\right) ,
\label{HB}
\end{equation}
where $\vec{\pi } = \vec{p} + e\vec{A}/c$, $\vec{p}$ is the electron momentum 
(two-dimensional), $\vec{A} = (0,Bx, 0)$ is the vector potential, and $\gamma$
is the band parameter. 

We evaluate the energy spectrum of an electron in a magnetic field and a 
periodic potential expressing the Hamiltonian (\ref{HV}) in the basis of 
eigenfunctions of the Hamiltonian (\ref{HB}). These eigenfunctions are specified
by the Landau index $n=0, \pm 1, \pm 2, \ldots$ and a parameter $k$, which is 
the $y$ component of the wave vector. The eigenfunctions of the Hamiltonian
(\ref{HB}) are given by
\begin{equation}
\Psi^{}_{n,k} = C^{}_n
\left( \begin{array}{c}
 {\rm sgn} (n) i^{|n|-1} \varphi^{}_{|n|-1,k} \\
    i^{|n|} \varphi ^{}_{|n|,k}
\end{array}  
 \right),
\label{f1}
\end{equation}
where $C^{}_n = 1 $ for $n=0$ and $C^{}_n = 1/\sqrt{2}$ for $n\neq 0$; ${\rm 
sgn} (n)=1$ for $n>0$, ${\rm sgn} (n)=0$ for $n=0$, and ${\rm sgn} (n)=-1$ for 
$n<0$. Here $\varphi^{}_{n,k}$ is the electron wave function with 
parabolic dispersion relation in the $n$-th LL
\begin{equation}
\varphi^{}_{n,k} (x,y) = \frac{e^{i k y}}{\sqrt{L}} 
\frac{e^{-(x-x^{}_k )^2/2\ell_0^2}}{\sqrt{ \pi^{1/2} \ell^{}_0 2^n n!}}
 H^{}_n (x-x^{}_k),
\label{PhiK}
\end{equation}
where $L$ is the length in the $y$ direction, $x^{}_k=k\ell_0^2$, 
$\ell^{}_0 = \sqrt{c\hbar/eB}$ is the magnetic length, and $H^{}_n(x)$ are the 
Hermite polynomials. The eigenenergy corresponding to the wave function (\ref{f1}) is 
$\varepsilon^{}_n = {\rm sgn}(n)\hbar \omega^{}_B \sqrt{|n|}$ \cite{aoki,abergeletal,ssc_13}.


\begin{figure}
\begin{center}\includegraphics[width=8cm]{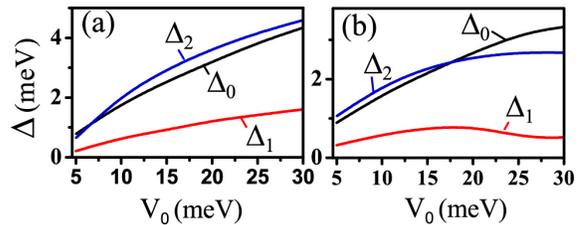}\end{center}
\caption{The band gaps in the $n=0$, $n=1$, and $n=2$ LLs versus
the amplitude of the periodic potential, $V^{}_0$, for interacting 
systems with half filling of the $n=0$ LL. The band gaps are 
defined as the gaps between the corresponding bands of Dirac fermions in a
magnetic field corresponding to $\alpha = 1/2$. The period of the potential is 
(a) $a^{}_0=20$ nm and (b) $a^{}_0 = 40$ nm. 
}
\label{fig2}
\end{figure}

We evaluate the matrix elements of the periodic potential $V(x,y)$ in the basis
$\Psi^{}_{n,k} (x,y)$ and construct the single-particle Hamiltonian matrix in 
the basis of the wave functions (\ref{f1}). In what follows, we consider only 
the basis states for $n=0$, $n=\pm 1$, and $n=\pm 2$ LLs of graphene. 
Inclusion of higher LL states does not change considerably the 
results for the band gaps presented below. From the Hamiltonian matrix we 
evaluate the energy spectrum of Dirac fermions in a magnetic field and a 
periodic potential. The corresponding spectra are shown in Fig.\ \ref{fig1} 
for $n=0$ and $n=\pm 1$ Landau levels. The energy spectra are shown as a function of 
the magnetic field in terms of the parameter $\alpha = \phi^{}_0/\phi$, where 
$\phi = Ba_0^2$ is the magnetic flux through the unit cell of the periodic 
potential and $\phi^{}_0 = h/e$ is the flux quantum. The energy spectra 
clearly show the fractal butterfly structure. For small values of $V^{}_0$ 
the coupling of the states of different LLs is small, and in each 
LL, for $\alpha= p/q$ with integer $p$ and $q$, there are $q$ bands. 
With increasing $V^{}_0$ the coupling of the states of different LLs 
becomes strong, which results in a strong overlap of different Landau bands. 

With the single-particle states at our disposal, we now calculate the matrix 
elements of the electron-electron (Coulomb) interaction and find the 
single-particle energy spectrum of the corresponding Hartree Hamiltonian. Since 
the electron density has the same periodic behavior as the periodic potential, 
the Fourier components of the Hartree potential are nonzero only for the discrete 
reciprocal vectors $\vec{G}$ and can be found from $V(\vec{G}) 
= (2\pi e^2/\kappa |\vec{G}|) n(\vec{G})$ and $V(0) = 0$. Here $n(\vec{G})$ is the 
Fourier component of the electron density that is determined by the occupied levels.  
For the many-particle system, we consider only $n=0$, $n=\pm 1$, and $n=\pm 2$. 
We also restrict the number of states in a given LL, i.e., we consider a 
finite size system with 5000 states per LL with the inter-wave vector separation
$\Delta k = q^{}_0/50$, where the one dimensional wave vector $k$ determines the electron 
state [see Eq.\ (\ref{PhiK})]. This corresponds to the size of the system in the 
real space to be $50a^{}_0 \times 50 a^{}_0$. To eliminate the boundary effects we consider 
the periodic boundary conditions. Due to the finite size of the system it is difficult to 
identify and trace the band structure of the energy spectra for generic rational values 
$p/q$ of $\alpha$. Therefore in what follows, the interaction effects 
on the band structure of graphene was studied for $\alpha=1/2$ and 
$\alpha =1/3$. For a noninteracting system at $\alpha=1/2$, in each LL 
there are two bands with zero band gap. In this case the inter-electron 
interaction opens a finite gap. For $\alpha = 1/3$, both noninteracting and interacting 
electron systems have three bands with two finite band gaps in each LL. 

\begin{figure}
\begin{center}\includegraphics[width=7cm]{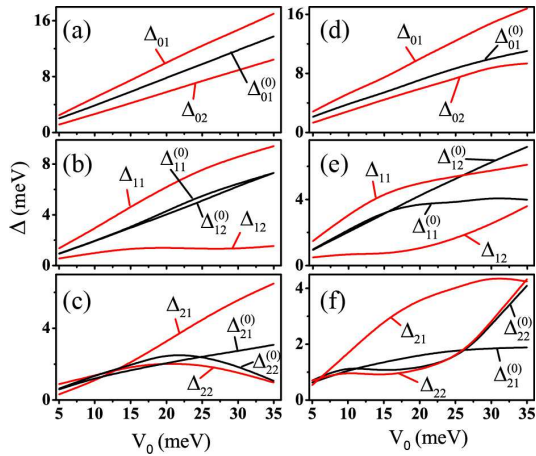}\end{center}
\caption{The band gaps versus $V^{}_0$ for $n=0$ (a,d), $n=1$ (b,e), and $n=2$ (c,f) 
LLs. The band gaps are defined as the gaps between the corresponding bands of Dirac 
fermions in a magnetic field for $\alpha = 1/3$. The black lines correspond to 
the case of the nonintercting system, while the red lines correspond to the Dirac fermions 
with Hartree interaction and half filling of the $n=0$ LL. The gaps are labeled 
as $\Delta_{ni}^{(0)}$ (noninteracting system) and $\Delta^{}_{ni}$ (interacting system), 
where $n$ is the LL index and $i=1$ and 2 corresponds to the low-energy 
and high energy gaps, respectively. The period of the periodic potential is 
20 nm (a,b,c) and 40 nm (d,e,f).
}
\label{fig3}
\end{figure}

The effects of interaction on the band structure also depend on the LL
filling. We present the main results for half filling of the $n=0$ LL, 
which is defined through zero Fermi energy. The single-particle spectra show 
the band structure with two bands in each LL. An increase in amplitude 
of the periodic potential causes both the band gap and band widths to increase in 
each LL. 

The influence of electron-electron interactions on the band gaps $\Delta ^{}_n$ for 
$\alpha =1/2$ is plotted in Fig.\ \ref{fig2} as a function of  $V^{}_0$. Here the band 
gap for the LL $n$ is labeled as $\Delta^{}_n$. Nonineracting systems have zero gaps
and are not shown here. The results are for $a^{}_0 = 20$ nm [Fig.\ \ref{fig2}(a)] and 
40 nm [Fig.\ \ref{fig2}(b)]. In general, in both cases the band gaps monotonically 
increase with $V^{}_0$. With the Landau level index the band gaps have a
strong nonmonotic dependence, where the band gap for $n=1$ is much 
smaller than those for $n=0$ and $n=2$ \cite{notes_2}. For the $n=1$ gap
there is also a small nonmonotonic dependence of $\Delta^{}_1$ on $V^{}_0$ with a 
local maximum for $V^{}_0\approx 17$ meV. 

\begin{figure}
\begin{center}\includegraphics[width=5cm]{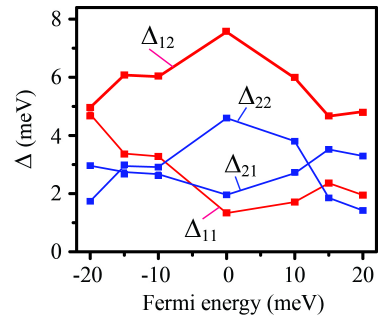}\end{center}
\caption{The band gaps for $n=1$ and $n=2$ LLs as a function of the Fermi energy
for electrons in $n=0$, i.e., as a function of the population of the $n=0$ LL
and for a magnetic field corresponding to $\alpha = 1/3$. The period and amplitude of
the potential are $a_0 = 20$ nm and $V_0 = 25$ meV, respectively. The zero Fermi
energy corresponds to half filling of the $n=0$ LL.
}
\label{fermi}
\end{figure}

The $\alpha=1/3$ results are shown in Fig.\ \ref{fig3}. In this case, both the 
noninteracting and interacting systems have three bands in each LL with 
two corresponding gaps. We label these gaps as $\Delta_{n,i}^{(0)}$ (noninteracting) 
and $\Delta^{}_{n,i}$ (interacting). Here $i=1,2$ is the number of the band gap: 
$i=1$ - low energy gap and $i=2$ - high energy gap. In the lowest LL 
[Fig.\ \ref{fig3}(a,d)], which is partially occupied, the gaps monotonically 
increase with $V^{}_0$. The interaction enhances the low-energy gap $\Delta^{}_{01}$ 
and suppresses the higher energy gap $\Delta^{}_{02}$. This suppression decreases 
with increasing period of the potential, $a^{}_0$. Similar behavior is also 
observed for $n=1$ [Fig.\ \ref{fig3}(b,c)], but now the suppression of the higher 
energy gap is large compared to the $n=0$ case. The gap $\Delta^{}_{12}$ becomes 
vanishingly small for low periods of the potential. Therefore the inter-electron 
interactions almost close one of the gaps in the energy dispersion for $n=1$. 
For $n=2$ the main effect of the interaction is a strong enhancement of the low-energy gap, 
$\Delta^{}_{21}$, while the higher energy gap is almost unaffected by the interaction.  
 
The general effects of the electron-electron interaction on the band structure of the 
energy dispersion is therefore an enhancement of the low energy gaps and suppression 
of the higher energy gaps (within a single LL). This suppression 
can be attributed to the inter-state repulsion introduced by the interaction
between the states of $n=0$ LL and the states of higher LLs.
The repulsion becomes weaker when the energy separation between the states increases. 
For $n=1$, which is the closest to the $n=0$ LL the effect of state repulsion 
is more pronounced which results in almost collapsing of the high-energy gap. The energy gaps 
(and the corresponding energy spectra) versus $V^{}_0$ approximately depend on 
$V^{}_0/\hbar\omega^{}_B\propto V^{}_0 a^{}_0$ (or $V^{}_0/(e^2/\ell^{}_0)\propto
V^{}_0 a^{}_0$). The minimum of $\Delta^{}_{22}$ in Fig.~3 (f), for example, is therefore also
visible in Fig.~3(c), but at higher values of $V^{}_0$.

The above results (Figs.~2 and 3) are for the half-filled $n=0$ LL, which corresponds 
to zero Fermi energy. Variation of the population in the $n=0$ LL, which can
be described numerically in terms of variation of the Fermi energy, also changes the gaps. As
an illustration of this dependence, in Fig.~4 the gaps $\Delta_{1,i}$ and $\Delta_{2,i}$ 
are shown as a function of the Fermi energy. This dependence reveals that the
difference between the gaps in the same LL (i.e., between $\Delta_{11}$ and $\Delta_{12}$)
is the largest for zero Fermi energy. Therefore, in this case we should expect the strongest
interaction effect on the energy spectra and the corresponding gaps.    

Both the periodic potential and the electron-electron interaction cause mixing 
of the states of different LLs. The typical energy scale of the Coulomb
interaction corresponding to the periodic potential with period $a^{}_0$ is $e^2/
\kappa a^{}_0\approx 20$ meV for $a^{}_0 = 20$ nm and the dielectric constant 
$\kappa = 4$. For the inter-Landau separation $\approx 50$ meV, this interaction 
strength results in a LL mixing. The periodic potential, which can be of 
the same order as the interaction strength, also introduces LL mixing, 
which increases with increasing $V^{}_0$. Since the magnetic field is proportional 
to $1/\alpha$ with decreasing $\alpha$ the inter-Landau level energy separation 
increases, which should suppress the inter-level mixing due to the inter-electron 
interactions and the periodic potential. 

\begin{figure}
\begin{center}\includegraphics[width=6cm]{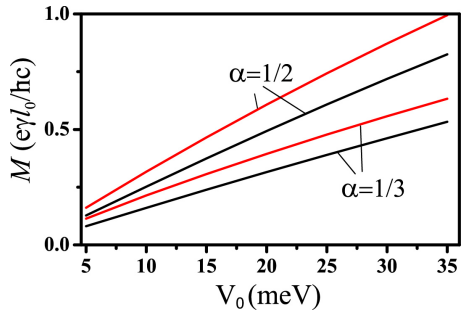}\end{center}
\caption{Magnetization of Dirac fermions as a function of the amplitude of
$V^{}_0$ for noninteracting (black line) and interacting 
(red line) systems and for two values of the parameter $\alpha$.  
The filling factor of the $n=0$ LL is $\nu = 1/2$.
}
\label{magnetization}
\end{figure}

A measurable quantity of the many-fermion system which depends on the LL 
mixing and can exhibit the effect of the electron-electron interaction is 
the magnetization $M = \frac{\partial {\cal H}}{\partial B}$ \cite{magnetization},
where the dependence of the Hamiltonian on the magnetic field is introduced 
through the vector potential $\vec{A}$. For the many-fermion system the 
magnetization operator is the sum of the single-particle contributions. The 
magnetization is then calculated as an expectation value of the magnetization 
operator. 

The magnetization of the non-interacting system and the system with 
electron-electron interactions are shown in Fig.\ \ref{magnetization}. For the 
non-interacting system the magnetization increases monotonically with increasing 
strength of $V^{}_0$. The reason for such an increase is 
the enhancement of the LL mixing with increasing $V^{}_0$. Without 
the LL mixing, the Dirac fermions in the $n=0$ LL have zero 
magnetization for all magnetic fields $B$, which follows from the fact that the 
energy of the $n=0$ LL is zero for all $B$ \cite{aoki,abergeletal,ssc_13}. 
Therefore, the nonzero magnetization illustrates the strength of the LL mixing. 
For the interacting system the magnetization monotonically increases with $V^{}_0$ 
and shows enhancement compared to the noninteracting system. With increasing $V^{}_0$ 
the mixing of the levels due to interaction increases both for $\alpha=1/2$ and 
$\alpha =1/3$. The magnetization increases with increasing $\alpha$, which 
illustrates a stronger inter-level mixing for larger values of $\alpha$. 

In conclusion, the interaction effects on the Hofstadter in monolayer graphene 
strongly depend on the amplitude of the periodic potential. For $\alpha = 1/2$, 
the interaction opens a gap in each LL, with the gap being smaller 
for $n=1$. For larger number of flux quanta per unit cell such as, $\alpha = 1/3$, 
the interaction suppresses the high-energy gaps and enhances the low-energy 
gaps compared to the noninteracting system. The effect is the strongest for 
$n=1$ where the high-energy gap is almost closed by the interaction. The 
magnetization of the system illustrates the enhancement of level mixing due 
to the interaction. This enhancement increases with increasing $V^{}_0$ and 
the level mixing becomes stronger with increasing $\alpha$. 

The work has been supported by the Canada Research Chairs Program of the 
Government of Canada.


\begin{thebibliography}{99}
\bibitem[\ddag]{byline} Electronic address:
Tapash.Chakraborty@umanitoba.ca

\bibitem{review}
U. R\"ossler and M. Shurke, in {\it Advances in Solid State Physics},
edited by B. Kramer (Springer, Berlin 2000), Vol. 40, pp. 35-50. 

\bibitem{harper}
P.G. Harper, Proc. Phys. Soc. London {\bf 68}, 874 (1955).

\bibitem{langbein}
D. Langbein, Phys. Rev. {\bf 180}, 633 (1969).

\bibitem{hofstadter}
D. Hofstadter, Phys. Rev. B {\bf 14}, 2239 (1976).

\bibitem{geisler_04}
M.C. Geisler, J.H. Smet, V. Umansky, K.von Klitzing, B. Naundorf, 
R. Ketzmerick, and H. Schweizer, Phys. Rev. Lett. {\bf 92}, 256801
(2004); Physica E {\bf 25}, 227 (2004).

\bibitem{albrecht_01}
C. Albrecht, J.H. Smet, K. von Klitzing, D. Weiss, V.Umansky, and
H. Schweitzer, Phys. Rev. Lett. {\bf 86}, 147 (2001); Physica E
{\bf 20}, 143 (2003).

\bibitem{ensslin_96}
T. Schl\"osser, K. Ensslin, J.P. Kotthaus, and M. Holland, Europhys. Lett.
{\bf 33}, 683 (1996); Semicond. Sci. Technol. {\bf 11}, 1582 (1996).

\bibitem{microwave}
U. Kuhl and H.-J. St\"ockmann, Phys. Rev. Lett. {\bf 80}, 3232 (1998).

\bibitem{optical_lattice}
M. Aidelsburger, M. Atala, M. Lohse, J.T. Barreiro, B. Paredes, and 
I. Bloch, Phys. Rev. Lett. {\bf 111}, 185301 (2013); H. Miyake, 
G.A. Siviloglu, C.J. Kennedy, W.C. Burton, and W. Ketterle, Phys.
Rev. Lett. {\bf 111}, 185302 (2013). 

\bibitem{david_PRL}
D.S.L. Abergel and T. Chakraborty, Phys. Rev. Lett. {\bf 102}, 056807 (2009).

\bibitem{aoki}
H. Aoki and M.S. Dresselhaus (Eds.), {\it Physics of Graphene} 
(Springer, New York 2014). 

\bibitem{abergeletal}
D.S.L. Abergel, V. Apalkov, J. Berashevich, K. Ziegler, and
T. Chakraborty, Adv. Phys. {\bf 59}, 261 (2010).

\bibitem{dean_13}
C.R. Dean, L. Wang, P. Maher, C. Forsythe, F. Ghahari, Y. Gao, J. Katoch,
M. Ishigami, P. Moon, M. Koshino, T. Taniguchi, K.Watanabe, K.L. Shepard, 
J.Hone, and P. Kim, Nature {\bf 497}, 598 (2013).

\bibitem{hunt_13}
B. Hunt, J.D. Sanchez-Yamagishi, A.F. Young, M. Yankowitz, B.J. LeRoy,
K. Watanabe, T. Taniguchi, P. Moon, M. Koshino, P. Jarillo-Herrero, 
and R.C. Ashoori, Science {\bf 340}, 1427 (2013).

\bibitem{geim_13}
L.A. Pomomarenko, R.V. Gorbachev, G.L. Yu, D.C. Elias, R. Jalil, A.A. Patel, 
A. Mishchenko, A.S. Mayorov, C.R. Woods, J.R. Wallbank, M. Mucha-Kruczynski, 
B.A. Piot, M. Potemski, I.V. Grigorieva, K.S. Novoselov, F. Guinea, V.I. Falko
and A.K. Geim, Nature {\bf 497}, 594 (2013).

\bibitem{moire_1}
R. Decker, Y. Wang, V.W. Brar, W. Regan, H.-Z. Tsai, Q. Wu, W. Gannett,
A. Zettl, and M.F. Crommie, Nano Lett. {\bf 11}, 2291 (2011).

\bibitem{moire_2}
J. Xue, J. Sanchez-Yamagishi, D. Bulmash, P. Jacquod, A. Deshpande, K. Watanabe,
T. Taniguchi, P. Jarillo-Herrero, and B.J. LeRoy, Nat. Mater. {\bf 10}, 282 
(2011).

\bibitem{note_1} Superlattice with a shorter period requires a larger magnetic
field to fill up the unit cell. The larger magnetic field values make for larger
Landau level separation in general. In graphene, the Landau level separation is
larger for low filling fractions than in conventional semiconductor systems due
to the linear dispersion relation. Further, the superlattice modulation
potential between graphene and the hexagonal boron nitride is $\sim$ 10 meV,
which is quite substantial. All of these attributes make the butterfly spectrum
more robust in graphene. P. Kim, private communications (2013).

\bibitem{rhim}
J.-W. Rhim and K. Park, Phys. Rev. B {\bf 86}, 235411 (2012).

\bibitem{nemec}
N. Nemec and G. Cuniberti, Phys. Rev. B {\bf 75}, 201404 (2007);
R. Bistritzer and A.H. MacDonald, Phys. Rev. B {\bf 84}, 035440 (2011).

\bibitem{vidar}
V. Gudmundsson and R.R. Gerhardts, Surf. Sci. {\bf 361-362}, 505 (1996);
Phys. Rev. B {\bf 52}, 16744 (1995); Phys. Rev. B {\bf 54}, 5223R (1996).

\bibitem{read}
A. Kol and N. Read, Phys. Rev. B {\bf 48}, 8890 (1993).

\bibitem{doh_salk}
H. Doh and S.H. Salk, Phys. Rev. B {\bf 57}, 1312 (1998).

\bibitem{ssc_13}
T. Chakraborty and V.M. Apalkov, Solid State Commun. {\bf 175 - 176}, 123 (2013).

\bibitem{mono_FQHE} V.M. Apalkov and T. Chakraborty, Phys. Rev. Lett.
{\bf 97}, 126801 (2006).

\bibitem{bi_FQHE}
V.M. Apalkov and T. Chakraborty, Phys. Rev. Lett. {\bf 105}, 036801 (2010); 
V. Apalkov and T. Chakraborty, Phys. Rev. Lett. {\bf 107}, 186803 (2011).

\bibitem{notes_2}
The reason for $\Delta _1<\Delta _2$ between the gaps is  
that the Coulomb interaction strength between the electrons in the $n=0$
and $n=1$ LLs is less by $\approx 3.5$ meV than that between 
the electrons in the $n=0$ and $n=2$ LLs. The interaction potential 
was evaluated for the wave vector $\approx 2\pi /a_0$. 

\bibitem{magnetization}
See, for example, J.P. Eisenstein, H.L. Stormer, V. Narayanamurti, A.Y. Cho, 
A.C. Gossard, and C.W. Tu, Phys. Rev. Lett. {\bf 55}, 875 (1985); I. Meinel, 
D. Grundler, S. Gargst\"adt-Franke, C. Heyn, and D. Heitmann, Appl. Phys. Lett. 
{\bf 70}, 3305 (1997); I. Meinel, T. Hengstmann, D. Grundler, D. Heitmann, 
W. Wegscheider and M. Bichler, Phys. Rev. Lett. {\bf 82}, 819 (1999); I. Meinel, 
D. Grundler, D. Heitmann, A. Manolescu, V. Gudmundsson, W. Wegscheider and M. Bichler, 
Phys. Rev. B {\bf 64}, 121306 (2001); M. Zhu, A. Usher, A.J. Matthews, A. Potts, 
M. Elliott, W.G. Herrenden-Harker, D.A. Ritchie, and M. Y. Simmons, Phys. Rev. B 
{\bf 67}, 155329 (2003). 

\end{thebibliography}
\end{document}